\documentclass[aps,twocolumn]{revtex4}
\usepackage{graphicx}
\usepackage{psfig}
\usepackage{epsfig}
\usepackage{latexsym}
\begin{document}

\title{Comment on ``Systematics of the Induced Magnetic Moments in 5d
Layers and the Violation of the Third Hund's Rule''}

\maketitle

Recently, Wilhelm \textit{et al.}~\cite{Wilhelm} reported x-ray
magnetic circular dichroism (XMCD) measurements at the W and Ir
$L_{2,3}$ edges of W/Fe and Ir/Fe multilayers. The $5d$ spin and
orbital magnetic moments, $\mu_S$ and $\mu_L$, of W and Ir were deduced
by applying the sum rules.
These moments are induced in the $5d$ band, through hybridization,
by the spin polarization of the Fe, which is reflected in their small values.
 For Ir,  $\mu_S$ and $\mu_L$
align both parallel to the Fe spin moment, while for W they align both
anti-parallel to Fe.
As a result, it was suggested that in
the case of W, with its less than half filled
$5d$ band, this implies a violation of Hund's third rule.
It seems a misnomer to call this a violation of Hund's rule,
since Hund's rules do not 
generally apply to induced moments, being only strictly valid for isolated atoms.
However, on the basis of their results for only two systems, the authors
further concluded: \textit{``This remarkable finding shows that the
induced magnetic behavior of $5d$ layers may be radically different
than that of impurities and alloys.''}  Here we {\it show}, on the
basis of a systematic set of calculations, that the behavior of 
$\mu_S$ and $\mu_L$ of $5d$ magnetic impurities in an Fe host
\textit{ is not radically different} than that of $5d$ interface layers in Fe
multilayers, proving that their conclusion, inferred on the basis of two examples, is 
not correct.

Calculations were performed for the electronic and magnetic properties
of periodic multilayers Fe/Z consisting of 5 monolayers (ML) of
Fe(100) and 3 ML of the $5d$ transition metal, Z, using 
the relativistic spin-polarized linear
muffin-tin orbitals (LMTO) method.~\cite{Ebert} 
Figure 1 includes, besides the present results for the $5d$ interface
layer, two different impurity calculations. These refer to
a non-self-consistent fully-relativistic calculation using a
self-consistent scalar-relativistic potential~\cite{Ebert2} and a
self-consistent fully-relativistic calculation.~\cite{Kornherr}  
%%%The results show that full self-consistency only slightly affects the
%%%behavior of $\mu_S$ and $\mu_L$ as a function of Z.
The  $\mu_S$ of $5d$ impurities in Fe \cite{Kornherr,Tyer}
is aligned anti-parallel for the first part of the $5d$ transition
metal series, i.e.\ for less than half filled band. From Os onwards 
$\mu_S$ of the $5d$ impurity and Fe are  parallel aligned. For
$5d$ impurities to obey Hund's third rule, $\mu_L$ should be positive
throughout the $5d$ series: this is not the case for Re, Os, and Ir. 

The curves of the calculated  $\mu_S$ and $\mu_L$ for the multilayers
show a similar behavior as for the impurities. $\mu_L$, being more
sensitive to the structure and chemical environment, switches between
parallel and anti-parallel at nearly the same location,  resulting in
a negative value for W, Re, and Os. Therefore, the relative alignment of 
$\mu_S$ and $\mu_L$ in the $5d$ metals
is the same as in the single impurity case, 
with the exception of W and Ir.

%Discrepancy with the measurements may arise due to experimental
%complications of XMCD for this particular system, not mentioned
%in Ref.~\cite{Wilhelm}, such as the non-negligible contribution
%of the magnetic-dipole term,
%$T_z$, in the spin-moment determination and the
%averaging over all layers of Z, whereas $\mu_S$ and $\mu_L$ 
%are most pronounced in the interface layer. 

%The present calculations agree well with the experimental data: W: experiment $(-0.18,-0.016)$,
%theory $(-0.11,-0.031)$ for $\mu_S$ and $\mu_L$ respectively;
%Ir: experiment $(0.20,0.019)$, theory $(0.26,0.003)$. It just so happens that in the vicinity 
%of W and Ir a sign change of $\mu_L$ occurs (Fig.1). 

The present calculations agree well with the experimental data which are 
for $\mu_S$ and $\mu_L$ $(-0.18,-0.016)$ and $(0.20,0.019)$ in W and Ir
respectively, while the theory gives $(-0.11,-0.031)$ and $(0.26,0.003)$.
It just so happens that in the vicinity 
of W and Ir a sign change of $\mu_L$ occurs (Fig.1). 
%Therefore it should come as no surprize that $\mu_L$ of W 
%can be positive or negative at the interface or for the impurtity respectively. 
%Our results indicate that mostly band filling
%determines  the trend of  $\mu_S$ and $\mu_L$ over the $5d$ series
%since a similar effect also occurs in the $4d$ series.~\cite{Tyer}
Thus it seems obvious why a systematic study should not be based on these two data points alone.
Our results indicate that mostly band filling,
and to a lesser extent geometrical effects,  
%of either single impurity or interface,
determine  the trend of  $\mu_S$ and $\mu_L$ over the $5d$ series.
This finding is corroborated by the fact that a similar trend also occurs for the $4d$ series.~\cite{Tyer}

\begin{figure}[h!]
\begin{center}
\includegraphics[height=2.45in,width=2.8in]{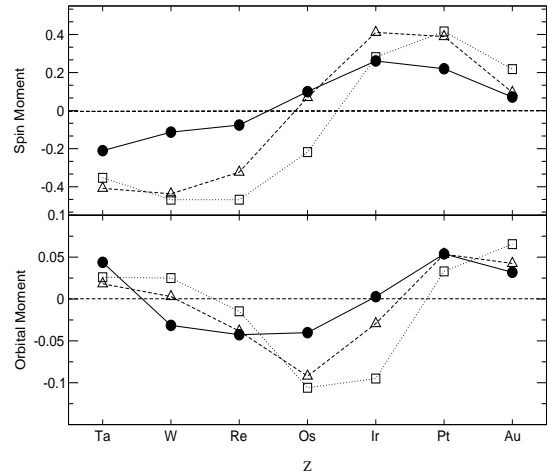}
\end{center}
\caption{Calculated spin magnetic moments ($\mu_S$, top panel)
and orbital magnetic moments ($\mu_L$, lower panel) in
$\mu_{\rm{B}}$/atom for $5d$ interface layers in
5Fe(100)/3Z multilayers, including interface relaxations, (Ref.~\cite{Tyer}:~$\bullet$) compared to
$5d$ impurities in Fe (Ref.~\cite{Ebert2}:~$\Box$ and
Ref.~\cite{Kornherr}:~$\triangle$)}
\end{figure}

\noindent
R.\ Tyer, G.\ van der Laan, W.M.\ Temmerman, Z.\ Szotek \\ 
Daresbury Laboratory \\
Daresbury, Warrington WA4 4AD, UK

%\smallskip \noindent
%H.\ Ebert \\
%University of Munich   \\              
%D-81377 Munich, Germany

\smallskip
Received 20 March 2002


%\end{thebibliography}

\end{document}